\documentclass[useAMS,usenatbib]{mn2e}
\usepackage{graphicx, rotating}
\usepackage{aas_macros}  
\usepackage{graphicx}
\usepackage{amssymb}
\usepackage{rotating}
\usepackage{multirow}
\usepackage[breaklinks,colorlinks=true,citecolor=black,linkcolor=black,urlcolor=blue,bookmarks=true]{hyperref}
\newcommand{\kms}{km\,s$^{-1}$}
\newcommand{\hi}{\textrm{H}\,\textsc{i}}
\newcommand{\hii}{\textrm{H}\,\textsc{ii}}
\newcommand{\ha}{\textrm{H}$\alpha$}
\newcommand{\msol}{\textrm{M}$_{\odot}$}

\title[Star formation history of UGCA\,92]{Star formation history and environment of
the dwarf galaxy UGCA\,92\thanks{Based
on observations made with the NASA/ESA Hubble Space Telescope,
obtained from the data archive at the Space Telescope Science Institute.
STScI is operated by the Association of Universities for Research
in Astronomy, Inc. under NASA contract NAS 5-26555.}}

\author[Makarova et al.]{
Lidia Makarova$^{1,2}$\thanks{E-mail: lidia@sao.ru},
Dmitry Makarov$^{1,2,3}$,
Sergey Savchenko$^{4}$\\
$^{1}$Special Astrophysical Observatory, Nizhniy Arkhyz, Karachai-Cherkessia 369167, Russia\\
$^{2}$Isaac Newton Institute of Chile, SAO Branch, Russia\\
$^{3}$Universit\'e Lyon~1, Villeurbanne, F-69622, France; CRAL, Observatoire de Lyon, St. Genis Laval, F-69561, France\\
$^{4}$Astronomical Institute, Saint-Petersburg State University, Saint-Petersburg, Russia
}

\begin{document}

\date{Accepted XXX. Received XXX; in original form XXX}

\pagerange{\pageref{firstpage}--\pageref{lastpage}} \pubyear{XXX}

\maketitle

\label{firstpage}

\begin{abstract}
We present a quantitative star formation history of the nearby dwarf galaxy UGCA\,92. This 
irregular dwarf is situated in the vicinity of the Local Group of galaxies in a zone of 
strong Galactic extinction (IC\,342 group of galaxies). The galaxy was resolved 
into stars with HST/ACS including old red giant
branch. We have constructed a model of the resolved stellar populations and measured the star 
formation rate and metallicity as function of time. 
The main star formation activity period occurred about 8 -- 14 Gyr ago. 
These stars are mostly metal-poor, with a mean metallicity [Fe/H] $\sim$
$-1.5$ -- $-2.0$ dex. About 84 per\,cent of the total stellar mass was formed during
this event. There are also indications of recent star formation starting about 1.5 Gyr ago 
and continuing to the present. The star formation in this event shows
moderate enhancement from $\sim$ 200 Myr to 300 Myr ago. It is very likely that
the ongoing star formation period has higher metallicity of about $-0.6$ -- $-0.3$ dex. 
UGCA\,92 is often considered to be the companion to the starburst
galaxy NGC\,1569. Comparing our star formation history of UGCA\,92 with that of
NGC\,1569 reveals no causal or temporal connection between recent star formation 
events in these two galaxies. We suggest that the starburst phenomenon in NGC\,1569 
is not related to the galaxy's closest dwarf neighbours and does not affect their 
star formation history. 
\end{abstract}


\section{Introduction}

\begin{figure*}
\includegraphics[height=8cm]{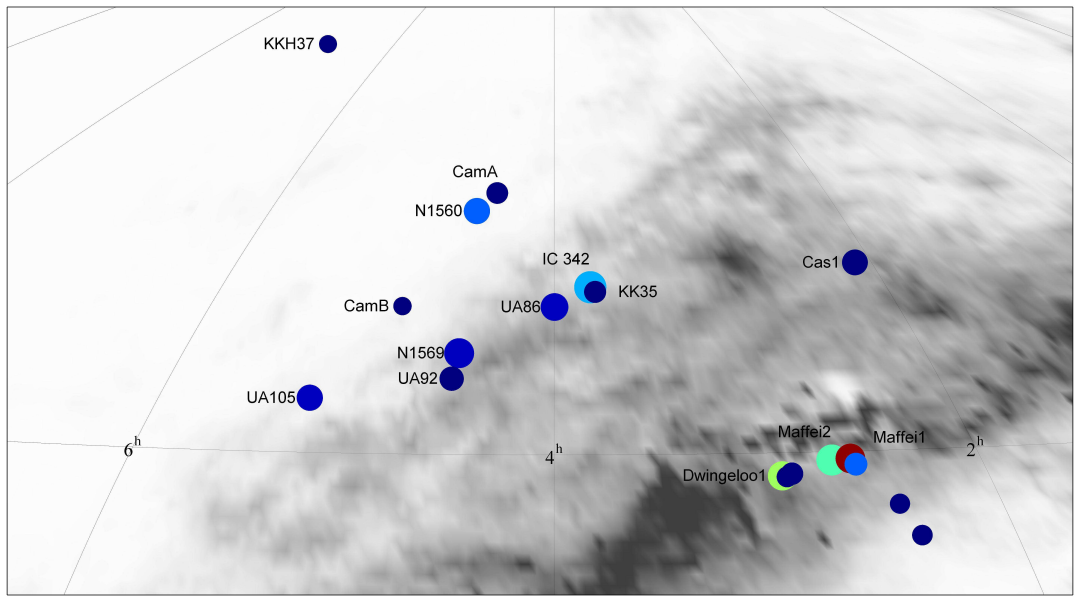}
\caption{Location of the IC\,342/Maffei complex of galaxies in Galactic coordinates.
Colours denote galaxy type. The \textit{IRAS} extinction map is shown in grey scale.}
\label{fig:dustgal}
\end{figure*}

Dwarf galaxies are most numerous objects in the Universe.
The question of star formation in dwarf galaxies is extremely important
for understanding of their origin and evolution.
The local Universe ($\leq$10 Mpc) is particularly important and convenient
for studying dwarf galaxies. 
Nearby galaxies are resolved into individual stars, 
which allow us to study stellar population of these galaxies directly.
In the last decades, significant progress
has been achieved in the study of resolved stellar populations due to 
the Hubble Space Telescope (\textit{HST}) and new large ground-based
telescopes.

About 50 per cent of nearby galaxies are situated in groups and clouds
\citep{mk2011}. 
Taking into account loose associations of dwarf galaxies
\citep{tully06}, most of the galaxies within 3 Mpc
are not isolated objects. 
The nearby group around IC\,342 is obscured by strong Galactic extinction
(see Fig.~\ref{fig:dustgal}). 
The IC\,342-Maffei complex should have significant impact on dynamic and evolution
of the nearby Universe. 
Unfortunately, the `Zone of Avoidance' hides this region from us and 
a determination of main properties of galaxies behind it is a challenge.

In the framework of the study of the structure of the nearby Universe (\textit{HST} 
project 9771)
we have obtained the images of dwarf galaxies within IC\,342/Maffei complex.
In the present work we have considered the star formation history of the dwarf
irregular galaxy UGCA\,92, which has for a long time been considered to be the closest
neighbour of the nearest starburst galaxy NGC\,1569 \citep{k1994,mk2003}.

The small irregular galaxy UGCA\,92 was discovered by \citet{nilson} and independently
catalogued as a possible planetary nebula by \citet{ell84}. CCD observations
of \citet{hoessel88} partially resolving it into individual stars in
the \textit{g} and \textit{r} passband images first showed that UGCA\,92 (EGB\,0427+63) 
is a dwarf irregular galaxy. \citet{hod95} detected 25 \hii{}
regions within UGCA\,92, concentrated in two well-separated regions
of the galaxy. 
The reddening, measured from the emission-line spectra for bright
\hii{} regions is $E(B-V)=0.90\pm0.08$ and the mean oxygen abundance
is about 13 per cent solar, with an uncertainty of 50 per cent. \citet{kar2010} carried out
\ha{} flux measurement for UGCA\,92 and derived a current star formation
rate of $\log(\textrm{SFR}) = -1.51$ \msol{}/yr.

The general parameters of UGCA\,92 are presented in Table~\ref{t:param}.
The total magnitudes, colours and central surface brightnesses, $\mu(0)$,
are not corrected for Galactic extinction.

\begin{table}
\caption{General parameters of UGCA\,92.}
\begin{tabular}{l@{}rl}
R.A.(J2000)                                 & $04^h32^m03.5^s$       & $[2]$ \\
Dec (J2000)                                 & $+63\degr36^\prime58^{\prime\prime}$          & $[2]$ \\[3pt]
size, arcmin                                & $2.0\times1.0$         & $[3]$  \\
Linear diameter, kpc                        & $3.1$                  & $[1]$ \\[3pt]
$(m-M)_0$, mag                              & $27.41\pm0.25$         & $[1]$ \\
Distance, Mpc                               & $3.03\pm0.35$          & $[1]$ \\[3pt]
$B_T$, mag                                  & $15.22$                & $[4]$ \\
$(B-V)_T$, mag                              & $1.34$                 & $[4]$ \\
$V_{3^{\prime}}$, mag                       & $14.55$                & $[5]$ \\
$I_{3^{\prime}}$, mag                       & $12.87$                & $[5]$ \\
$J_T$, mag                                  & $12.38$                & $[6]$ \\
$K\!s_T$, mag                               & $11.12$                & $[6]$ \\[3pt]
$\mu(0)_B$, mag\,arcsec$^{-2}$              & $25.1$                 & $[4]$ \\
$\mu(0)_V$, mag\,arcsec$^{-2}$              & $24.18\pm0.01$         & $[5]$ \\
$\mu(0)_I$, mag\,arcsec$^{-2}$              & $22.61\pm0.01$         & $[5]$ \\[3pt]
$E(B-V)$, mag                               & $0.79\pm0.13$          & $[7]$ \\
$A_I$, mag                                  & $1.54\pm0.25$          & $[7]$ \\[3pt]
$V_{LG}$, \kms{}                            & $93$                   & $[8]$  \\
$M_B$, mag                                  & $-15.61$               & $[1]$  \\
$M_{HI}$, \msol{}                           & $1.56\times10^8$       & $[8]$ \\
$M_{HI}/L_B$                                & $0.55$                 & $[8]$ \\[3pt]
Fraction of old stars (12--14 Gyr), \%      & $84\pm7$               & $[1]$ \\
Metallicity of old stars, [Fe/H], dex       & $-2.0$\,--\,$-1.5$     & $[1]$ \\
Fraction of young stars (500 Myr), \%       & $7.6\pm0.7$            & $[1]$ \\
$\langle\textrm{SFR}\rangle$, 12 -- 14 Gyr ago, \msol\,yr$^{-1}$ & $1.2\pm0.1\times10^{-1}$   & $[1]$ \\
$\langle\textrm{SFR}\rangle$, last 500 Myr, \msol\,yr$^{-1}$     & $4.3\pm0.6\times10^{-2}$   & $[1]$ \\
\hline
\multicolumn{3}{p{0.45\textwidth}}{
$[1]$~this work;\hspace{2mm}
$[2]$~LEDA;\hspace{2mm}
$[3]$~\citet{CNG};\hspace{2mm}
$[4]$~\citet{kar96};\hspace{2mm}
$[5]$~\citet{sharina};\hspace{2mm}
$[6]$~\citet{vaduv05};\hspace{2mm}
$[7]$~\citet{dustmap};\hspace{2mm}
$[8]$~\citet{figgs}
} 
\end{tabular}
\label{t:param}
\end{table}

\section{The Dataset}

\begin{figure*}
\includegraphics[height=12cm,clip]{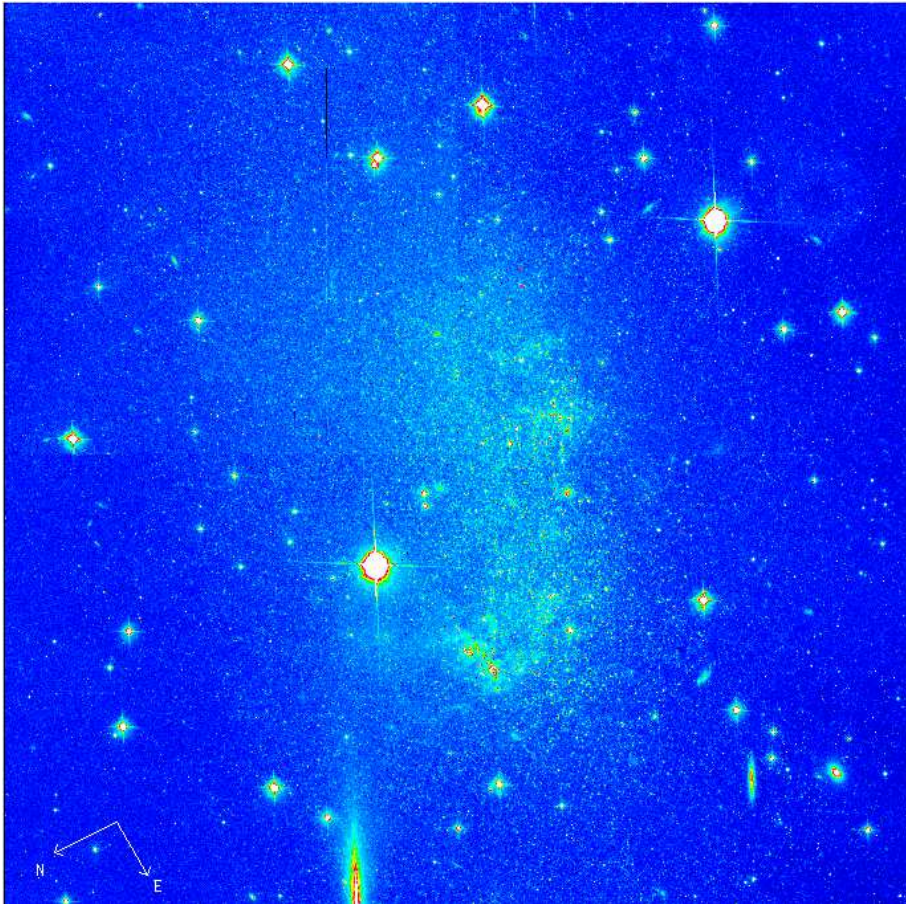}
\caption{\textit{HST}/ACS image of UGCA\,92 in \textit{F606W} filter. 
The image size is $3.4\times3.4$ arcmin.}
\label{fig:ima}
\end{figure*}

\begin{figure}
\centerline{\includegraphics[height=12cm,clip]{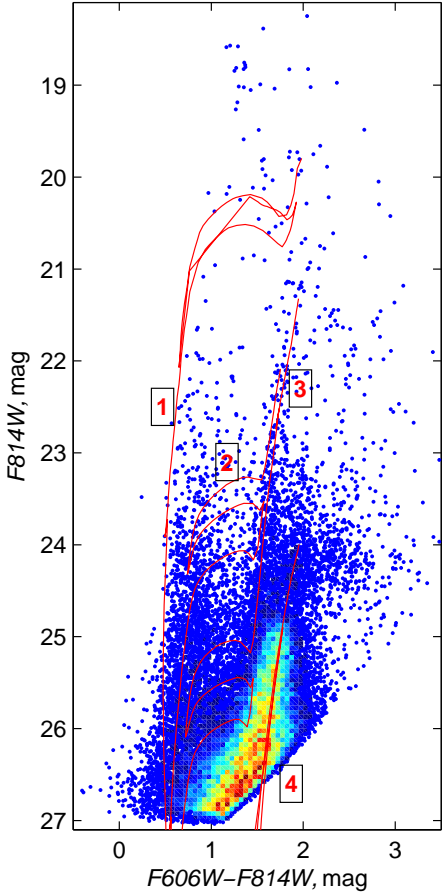}}
\caption{The $(\textit{F606W}-\textit{F814W})$, \textit{F814W} CMD of the dwarf galaxy UGCA\,92.
In the dense parts of the diagram the colour represents the density in the Hess diagram,
while individual stars are represented where they can be individually distinguished.
The magnitudes are not corrected for Galactic extinction.
Padova isochrones \citep{girardi00} corresponding to the mean
age and metallicity of detected star formation episodes are shown:
`1' is Z=0.008,t=10 Myr; `2' -- Z=0.001,t=50 Myr; `3' -- Z=0.0004,t=150 Myr;
`4' -- Z=0.0004,t=13 Gyr.}
\label{fig:cmd}
\end{figure}

\begin{figure*}
\includegraphics[width=8cm]{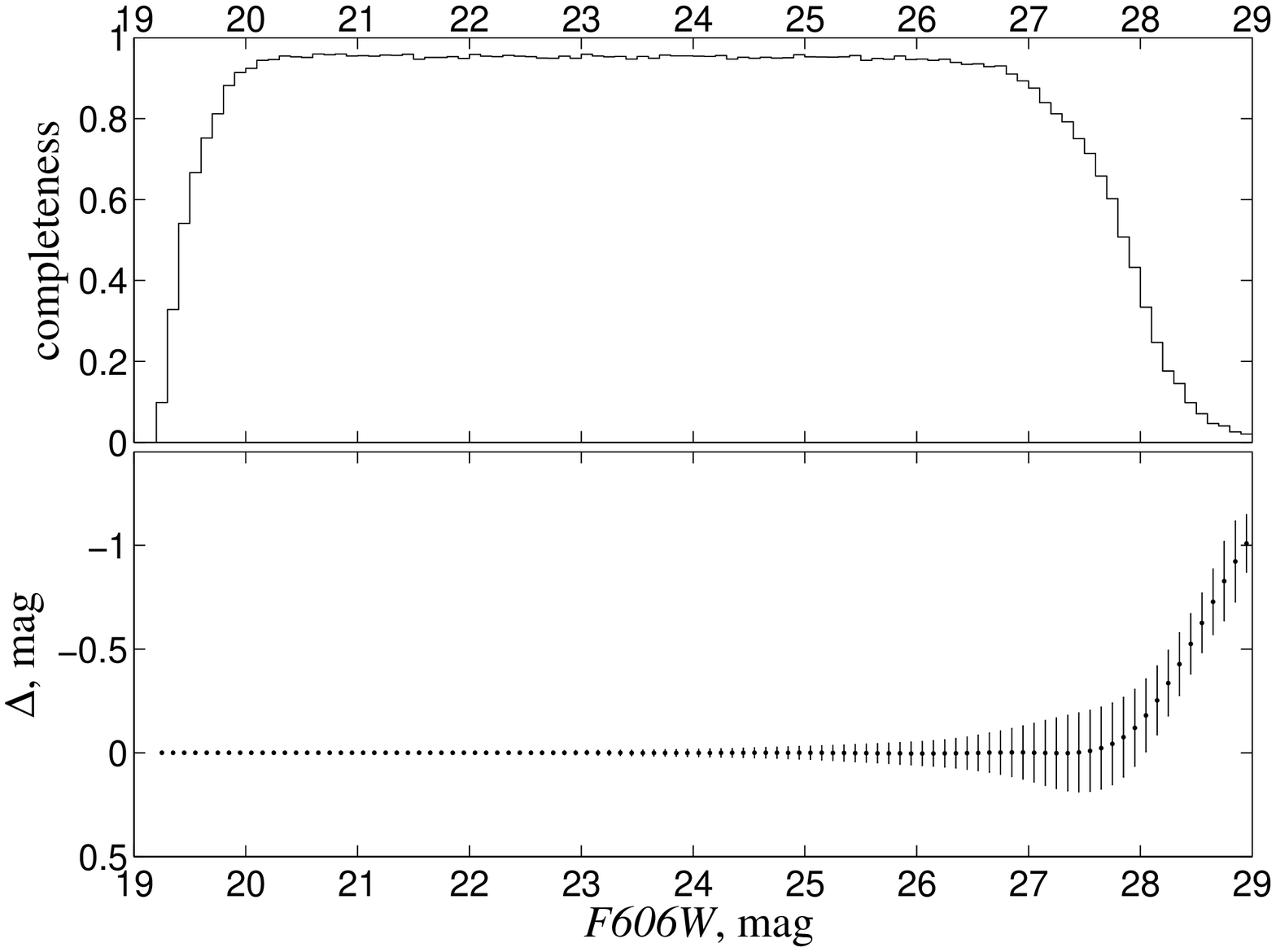}
\includegraphics[width=8cm]{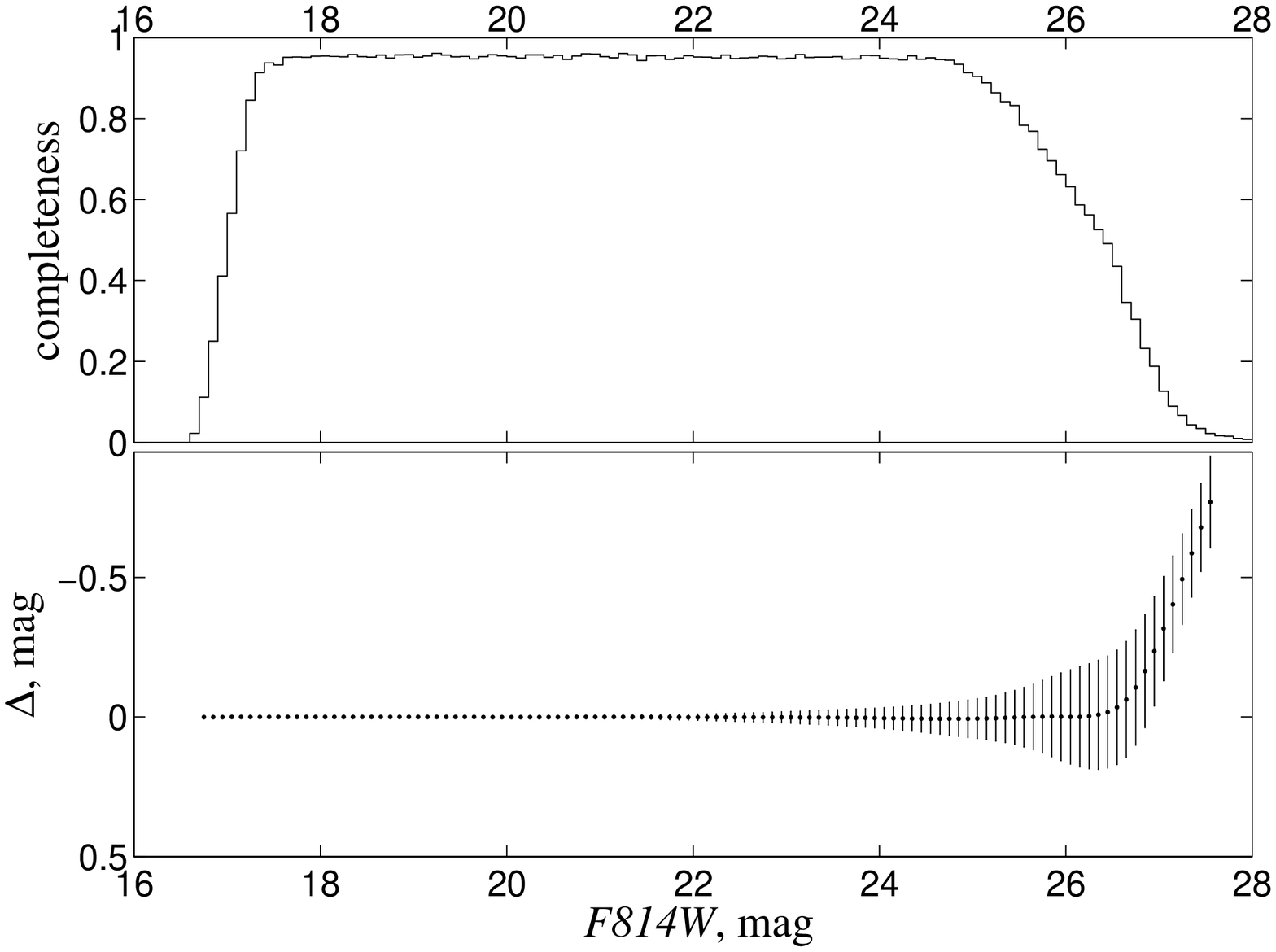}
\caption{Photometric errors and completeness for UGCA\,92.
The top panels show the completeness, i.e.\ the fraction of artificial stars
recovered within the photometric reduction procedure, as a function of
the \textit{F606W} and \textit{F814W} magnitudes.
The bottom panels give the difference between the measured and the true input
magnitude ($\Delta$mag = measure $-$ input). The error bars are 1\,$\sigma$
residuals.}
\label{fig:uncert}
\end{figure*}

The dwarf irregular galaxy UGCA\,92 was observed aboard \textit{HST} using Advanced Camera
for Surveys (ACS) at March 28, 2004 (SNAP 9771, PI I. Karachentsev).
Two exposures were made with the filters \textit{F606W} (1200 s) and \textit{F814W} (900 s).
An ACS image of the dwarf galaxy is shown in Fig.~\ref{fig:ima}.
The photometry of resolved stars in the galaxy was performed with the ACS module
of the \textsc{dolphot} package\footnote{\url{http://purcell.as.arizona.edu/dolphot/}}
for crowded field photometry \citep{dolphin02} using the recommended recipe and
parameters. Only stars with photometry of good quality were included in
the final list, following recommendations given in the \textsc{dolphot} User's Guide.
We have selected stars with signal-to-noise (S/N) $\ge 5$ in both filters,
$\chi^2\le2.5$ and $\vert \textit{sharp} \vert \le 0.3$. The resulting colour-magnitude
diagram contains 32699 stars (Fig.~\ref{fig:cmd}).

Artificial star tests were performed using the same reduction procedures
to estimate photometric errors, crowdedness and blending effects in
the most accurate way. A large library of artificial stars was generated spanning
the necessary range of stellar magnitudes and colours so that the distribution
of the recovered photometry is adequately sampled.
The photometric errors and completeness are presented in Fig.~\ref{fig:uncert}.
The 1\,$\sigma$ photometric precision is 0.07 at  $\textit{F814W}=25$ and
0.19 at $\textit{F814W}=27$ mag. 
Malmquist bias becomes notable for stars with $\textit{F814W} > 26.8$ mag. 
In \textit{F606W} the 1\,$\sigma$ photometric precision
is 0.06 at 26 and 0.18 at 28 mag.

\section{Colour-magnitude diagram}

All resolved stars are significantly shifted to redder colours due
to high extinction in the Zone of Avoidance of the Milky Way. The diagram is
typical for dwarf irregular galaxies. A pronounced upper main sequence (MS)
and probable helium-burning blue loop stars are found at
$\textit{F606W}-\textit{F814W} < 1.2$ mag. 
The red supergiant branch (RSG) and the intermediate age asymptotic giant branch
(AGB) are also well populated. The most abundant feature in the CMD is
the red giant branch (RGB) (see Fig.~\ref{fig:cmd}).

\subsection{Extinction}

The dwarf galaxy UGCA\,92 is situated at the Galactic latitude $l = +10.5^{\circ}$.
Galactic extinction in the field of UGCA\,92 needs to be addressed
in more detail
because reddening estimation at low Galactic latitude in the Zone of Avoidance 
is highly uncertain \citep[see][appendix C]{dustmap}.
\citet{dustmap} give a
colour excess of $E(B-V)=0.79\pm0.13$ using IRAS/DIRBE maps of
infrared dust emission. We apply this value to the colour-magnitude diagram and
estimate a mean colour of the upper MS as a first test of
the feasibility of the colour excess value. The unreddened $V-I$ colour of upper
MS stars is about zero \citep[see][for example]{kenyon}. We have selected
main sequence stars in the appropriate magnitude and colour range,
$\vert (V-I)_0 \vert \le 0.5$ and $19.5 \leq I_0 \leq 24.0$.
According to our measurements, the mean MS colour is $(V-I)_0 = 0.04$ mag.
The colour spread of the MS $\Delta(V-I)_0=0.20$ mag is rather large.
This spread, as well as the slightly reddish mean MS colour, could
indicate a presence of a blue loop population. Thereby, the extinction
value given by \citet{dustmap} seems reasonable, and we also do not expect
a significant value of internal extinction in this faint dwarf galaxy.

The reddening in UGCA\,92 was also determined by \citet{hod95} from
the spectroscopy of 25 bright \hii{} regions within the galaxy. Their
reddening value $E(B-V) = 0.90\pm0.08$ is in agreement with the value of
\citet{dustmap} within uncertainties.
The resolution in the IRAS/DIRBE maps is 6.1 arcmin \citep{dustmap}. It is considerably larger than the ACS field of view. Therefore, we cannot get an information about possible variations in extinction across the field of view from these maps. However, a simple test was made using a colour-magnitude diagram of UGCA\,92. A value of the tip of the red giant branch was measured in four different subframes along {\it y} axis of the whole ACS frame. We have found, that the TRGB value has no variations within 1.5 $\sigma$ of the TRGB uncertainty. Consequently, we suggest no variations of external extinction within the ACS field of UGCA\,92.

Therefore, we use the colour excess $E(B-V)=0.79\pm0.13$ from \citet{dustmap} in all our measurements in the present paper.

\subsection{Foreground contamination}

The colour-magnitude diagram is highly contaminated by Milky Way (MW) stars.
To account for this contamination we need to construct
colour-magnitude diagram of the MW in the direction of UGCA\,92. This
information can be obtained from images of neighbouring `empty' fields.
Three fields nearby UGCA\,92 were found in the \textit{HST} data archive exposed
with WFPC2 in the parallel mode (coordinates of the centre are $4^h29^m53^s$,
$+64\degr42^{\prime}34^{\prime\prime}$). However, the photometric limit of these
images is about 3 mag higher than photometric limit of the working UGCA\,92
images. To account for the MW contamination in the region of faint stars, we
have used the \textsc{trilegal} program \citep{trilegal}, which computes synthetic
colour-magnitude diagrams for the specified coordinates in the sky and given
parameters of the Milky Way models. Consequently, we use neighbouring field
stars to account for the Milky Way contribution to the UGCA\,92 CMD brighter
than $\textit{F814W}=24$ mag and the \textsc{trilegal} synthetic data for the fainter part of the CMD.
Thirty synthetic CMDs were constructed with \textsc{trilegal} and then averaged
to avoid stochastic errors in synthetic CMDs. Random and systematic photometric
uncertainties and completeness measured from artificial star experiments were
applied to the synthetic CMDs. Resulting contamination by the MW stars was
determined to be 202 stars over the CMD of UGCA\,92.

\section{Distance determination}

\begin{figure*}
\includegraphics[height=12cm,angle=-90]{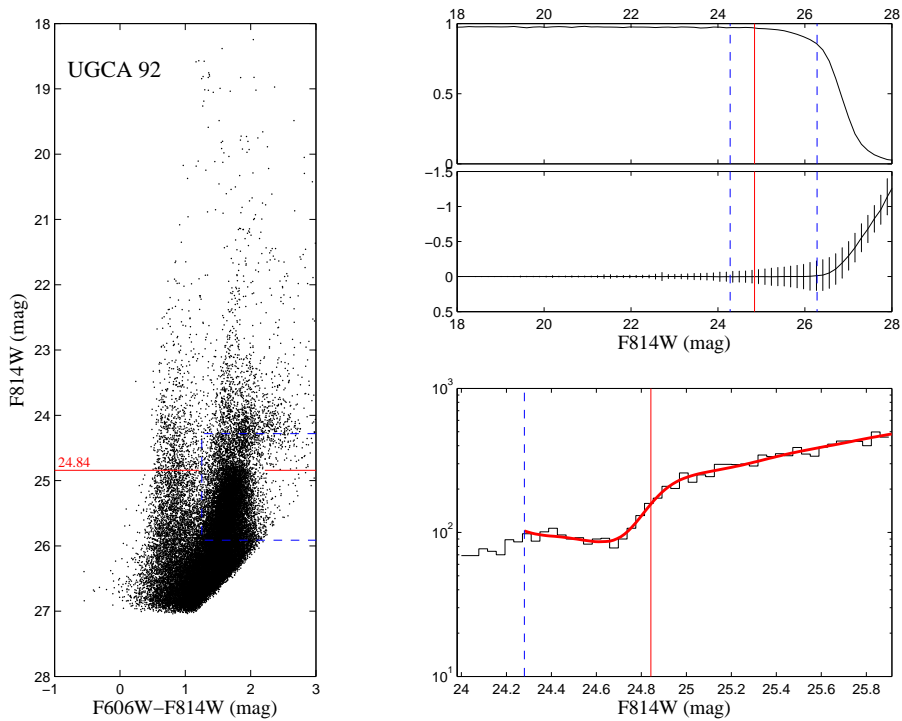}
\caption{ The colour-magnitude diagram (left panel) and the TRGB calculation results
for UGCA\,92. The upper right panel shows the completeness, photometric errors
and dispersion in errors (vertical bars) versus the \textit{HST}/ACS \textit{F814W} filter magnitude.
The lower right panel gives a histogram of the \textit{F814W} luminosity function.
The resulting model LF convolved with photometric errors and incompleteness
is displayed as a bold solid line with a jump at the position of the TRGB.}
\label{fig:trgb}
\end{figure*}

The distance of UGCA\,92 was first estimated by \citet{kts}
using the brightest blue and red supergiants as distance indicators.
The estimated distance modulus $(m-M)_0 = 26.72$ mag suffered from large
photometric uncertainties associated with crowded stellar fields and
uncertain reddening. Later \citet{karach97} estimated a
distance to UGCA\,92 using the same distance indicators and better
quality data. The resulting distance modulus is $(m-M)_0 = 26.25$ mag. 
Recently, more precise distance determination was made using \textit{HST}/ACS data
\citep{ketal06}. A deep colour-magnitude diagram permitted the use of the TRGB
distance indicator, resulting in a distance modulus of
$(m-M)_0 = 27.39$ mag. 

However, the TRGB method has recently been considerably improved.
We have determined the photometric TRGB distance with
our \textsc{trgbtool} program, which uses a maximum-likelihood algorithm
to determine the magnitude of the tip of the red giant branch from the stellar
luminosity function \citep{TRGB1}. The estimated value of TRGB is equal to
$\textit{F814W} = 24.84\pm0.02$ mag in the ACS instrumental system. The calibration
of the TRGB distance indicator has also recently been improved \citep{TRGB2}, 
where the colour dependence of the absolute magnitude of
the TRGB and zero-point issues in \textit{HST}/ACS and WFPC2 have been addressed. 
 Using this calibration, we have obtained the true distance modulus
to UGCA\,92 of $(m-M)_0 = 27.41\pm0.25$ mag and a distance of
$D = 3.03\pm0.35$ Mpc. 
Bear in mind that the given small error of the TRGB measurement (0.02 mag) and 
the high precision of the calibration (0.02 mag), the resulting accuracy is entirely 
determined by uncertainty of foreground extinction of 0.25 mag in the direction of UGCA\,92.
The colour-magnitude diagram with the fitted RGB luminosity
function and the resulting TRGB value is shown in Fig.~\ref{fig:trgb}.

\section{Star formation history}

The star-formation and metal-enrichment history of UGCA\,92 has been
determined from its CMD using our \textsc{StarProbe} package.
This program adjusts the observed photometric distribution of stars in the
colour-magnitude diagram against a positive linear combination
of synthetic diagrams of single stellar populations (SSPs, single age
and single metallicity). Our approach is described in more detail
in \citet{mm04} and \citet{mm10}.

\begin{figure*}
\includegraphics[width=12cm]{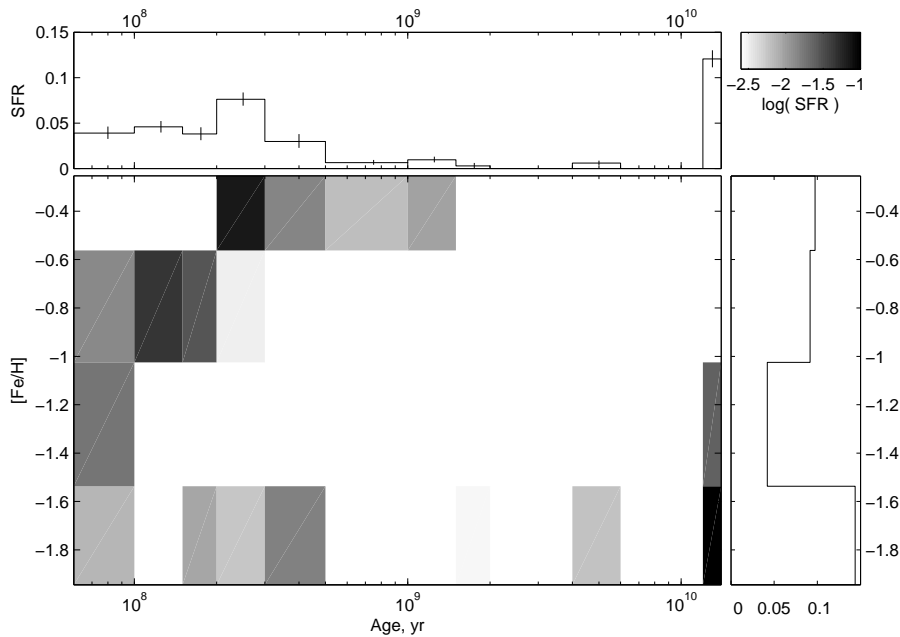}
\caption{The star formation history of the dwarf irregular galaxy UGCA\,92.
Top panel shows the star formation rate (SFR) ($M_\odot$/yr) against the
age of the stellar populations.
The bottom panel represents the metallicity of stellar content as function of
age. The colour corresponds to the strength of SFR for given age and
metallicity.}
\label{fig:cmd_sfh}
\end{figure*}

The observed data were binned into Hess diagrams, giving the
number of stars in cells of the CMD (two-dimensional histogram).
The size of the cell is 0.05\,mag in each passband.
The synthetic Hess diagrams were constructed from theoretical stellar
isochrones and initial mass function (IMF).
Each artificial diagram is a map of probabilities to find a star in a cell
for given age and metallicity.
We used the Padova2000 set of theoretical isochrones \citep{girardi00}, and a
\citet{salpeter55} IMF. The distance is adopted from the present paper
(see above) and the Galactic extinction from \citet{dustmap}.
The synthetic diagrams were altered by the same
incompleteness and crowding effects, and the photometric systematics,
as those determined for the observations using artificial stars
experiments. We also have taken into account the presence of unresolved
binary stars (binary fraction). Following \citet{barmina}, the binary
fraction was taken to be 30 per\,cent. The mass distribution for the secondary
was taken to be flat in the range 0.7 to 1.0 of the primary mass.
The best-fitting combination of synthetic CMDs is a maximum-likelihood solution
taking into account the Poisson noise of star counts in the cells of the Hess diagram.
The resulting star formation history (SFH) is shown in the Fig.~\ref{fig:cmd_sfh}.
The 1\,$\sigma$ error of each SSP is derived from an analysis of likelihood function.

According to our measurements, the main star formation event occurred
in the period 12 -- 14 Gyr ago with a rather high mean star formation rate (SFR)
of $1.2\pm0.1\times10^{-1}$ \msol{} yr$^{-1}$. This is the total SFR over
the whole galaxy. The metallicity range for the most stars formed during this event
is ${\rm [Fe/H]}=[-2.0:-1.5]$\,dex. This initial burst accounts for
about 84$\pm7$ per\,cent of the total mass of formed stars.

The quiescence period has appearing from about 6 to 12 Gyr ago. However,
the absence of star formation activity in this period
could be due to tight packing of the different age stars in the upper part of
the RGB. At the distance of 3 Mpc, fainter stellar populations,
like a horizontal branch and a lower part of the main sequence,
are hardly resolved at the \textit{HST}. Without these details it is difficult to resolve
an age-metallicity-SFR relation for the oldest ( $>6\div8$ Gyr) star formation
events, due to tight packing of the correspondent isochrones for the
brightest part of the CMD.

There are signs of marginal (insignificant) star formation 4 -- 6 Gyr ago. A
metallicity of these stars is similar to the metal abundance of the oldest
stellar population.

There are also indications of recent star formation starting about 
1.5 -- 2 Gyr ago and continuing to the present. 
We have measured the SFH in short age periods in the last 500 Myr 
to give more detail for the recent and ongoing star formation.
A mean SFR of the stars formed in the last 50 Myr is
$3.9\pm0.6\times10^{-2}$ \msol{} yr$^{-1}$.
This ongoing star formation rate is in good agreement with the independent
estimation by \citet{kar2010} from \ha{} flux measurements
($3.2\pm0.3\times10^{-2}$ \msol{} yr$^{-1}$).
The recent star formation is showing moderate enhancement from $\sim$ 200 Myr to
300 Myr. A mean star formation rate in the last 500 Myr is 
$4.3\pm0.6\times10^{-2}$ \msol{} yr$^{-1}$.
A mass portion of stars formed in the last 500 Myr
is 7.6$\pm0.7$ per\,cent of the total stellar mass.
A metallicity of the recent star formation event is determining with
large uncertainty due to relatively poor statistic in comparison with
sufficiently more numerous old stars. However, the measurements show that
a significant part of the young stars is evidently metal enriched.
It is very likely that the ongoing star formation has a metallicity of
$-0.6$ -- $-0.3$ dex.

\subsection{Spatial distribution of stellar populations}

\begin{figure*}
\includegraphics[width=8cm,angle=-90,clip]{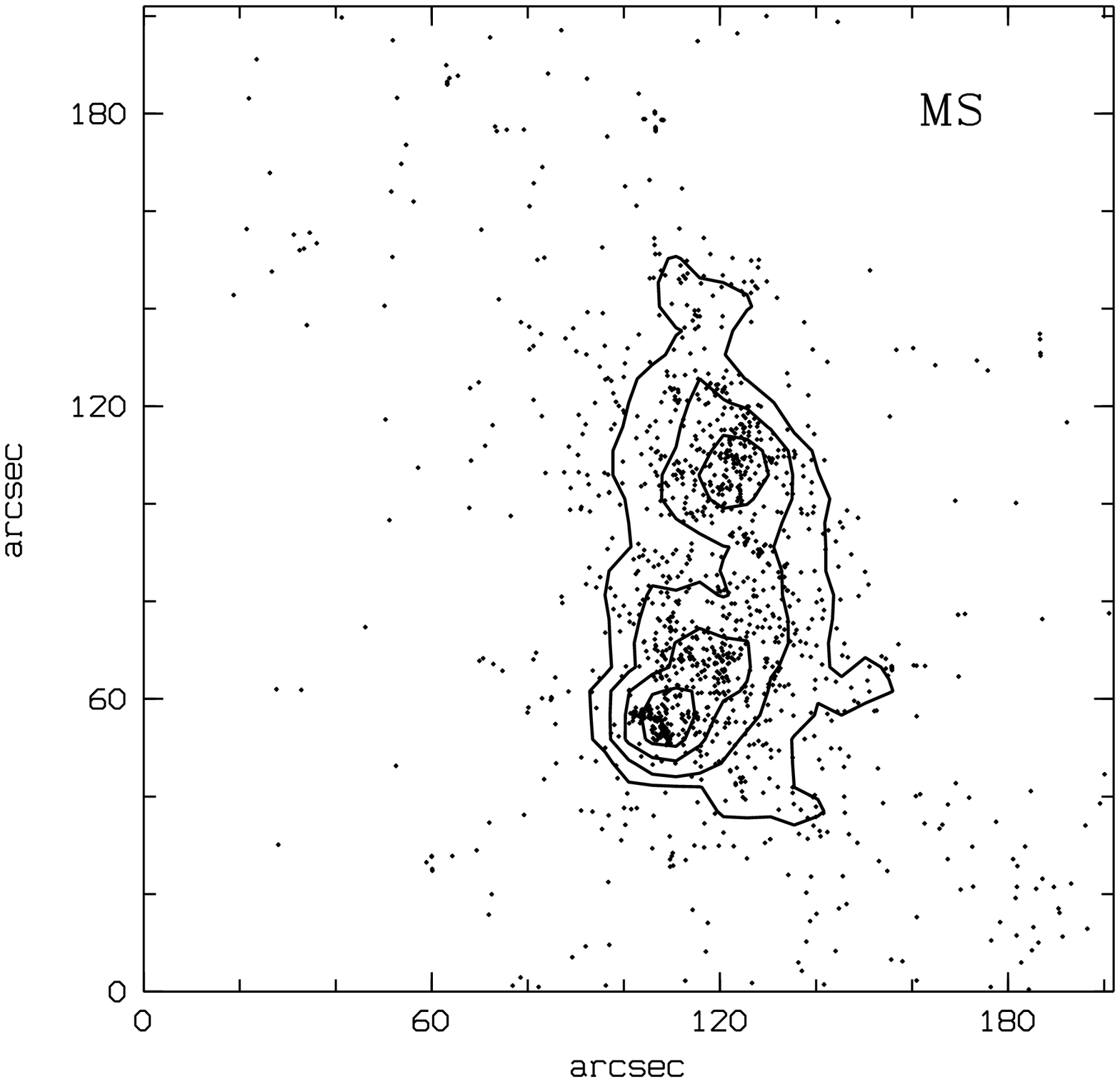}
\includegraphics[width=8cm,angle=-90,clip]{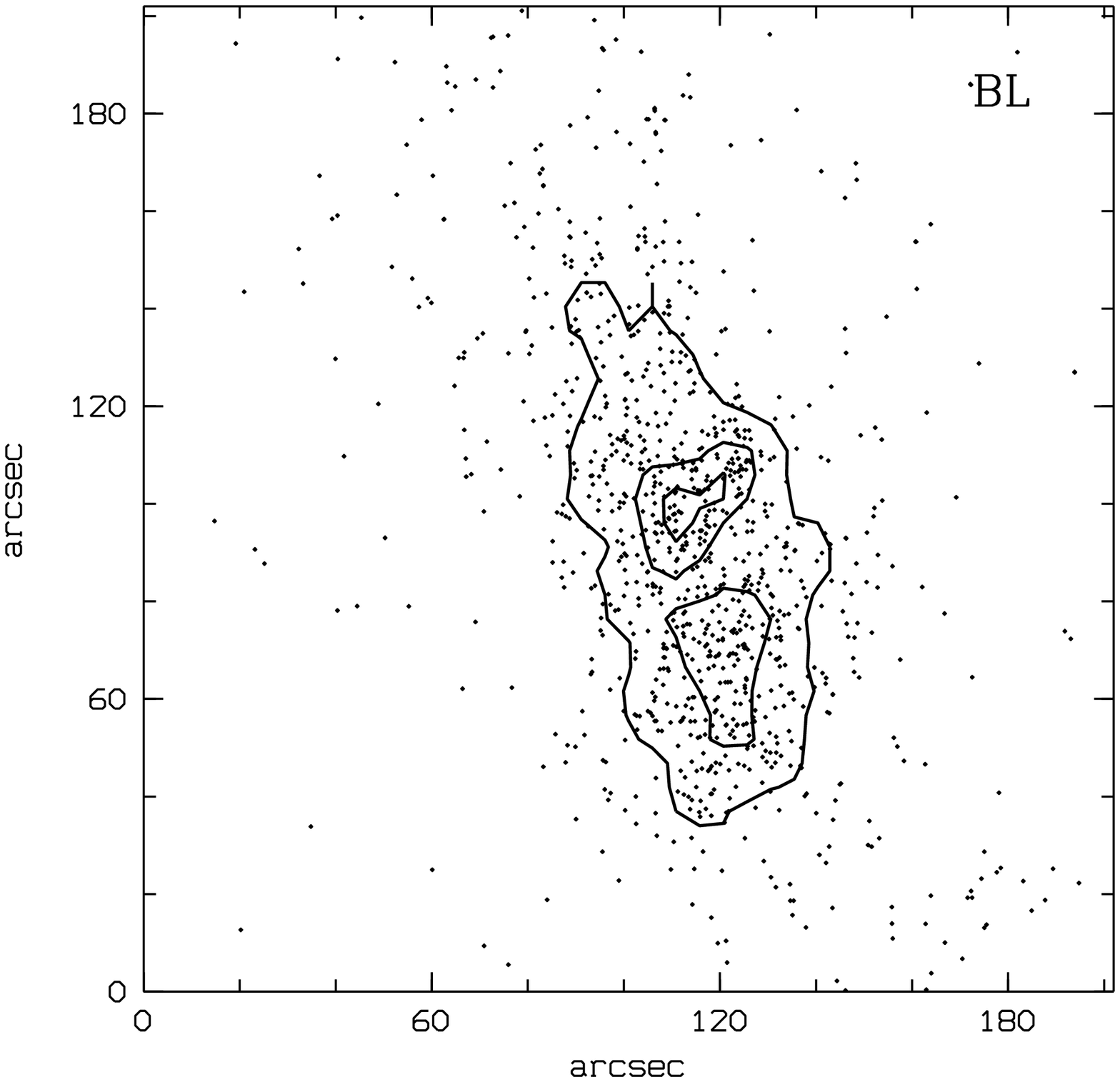}
\includegraphics[width=8cm,angle=-90,clip]{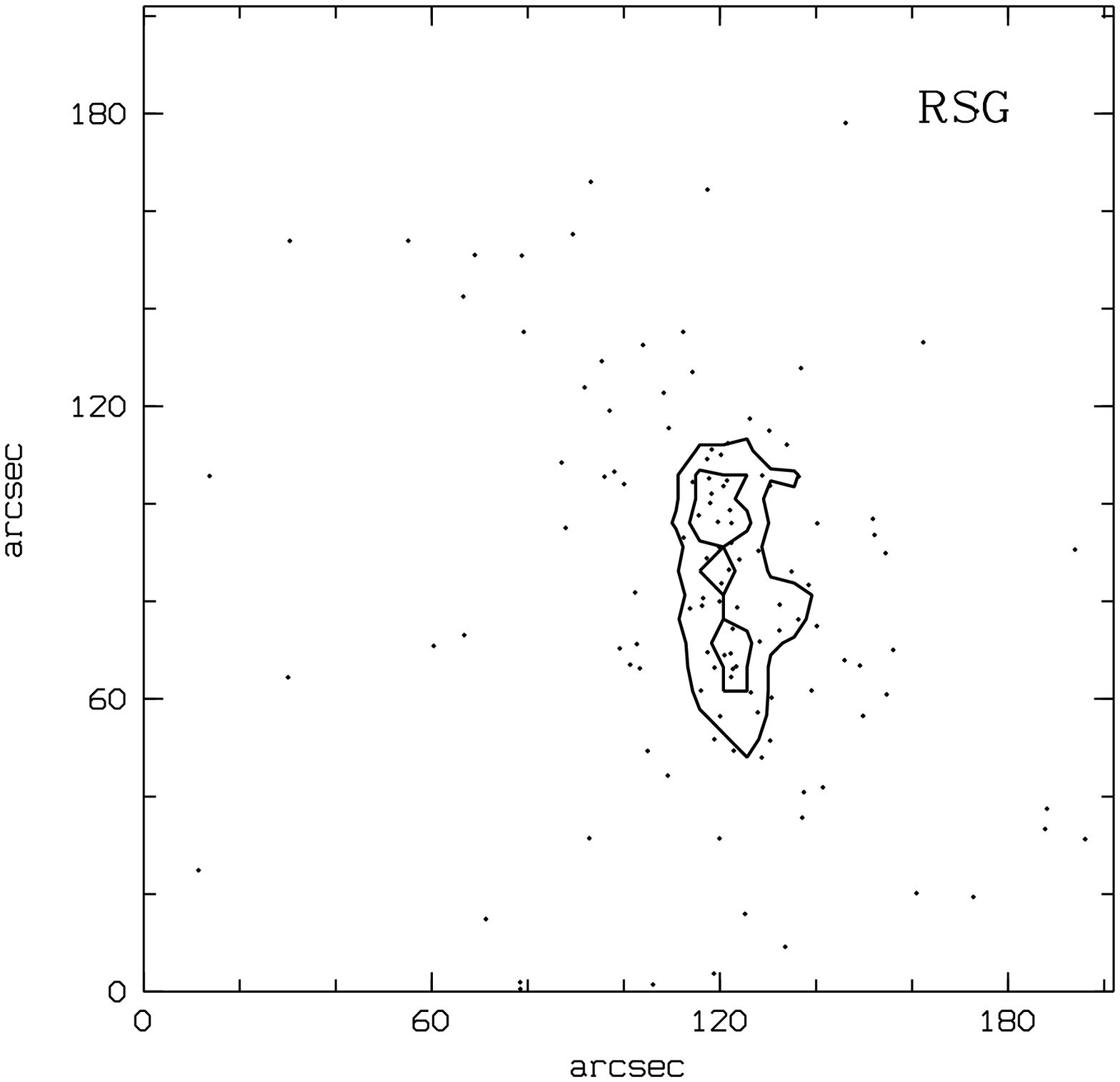}
\includegraphics[width=8cm,angle=-90,clip]{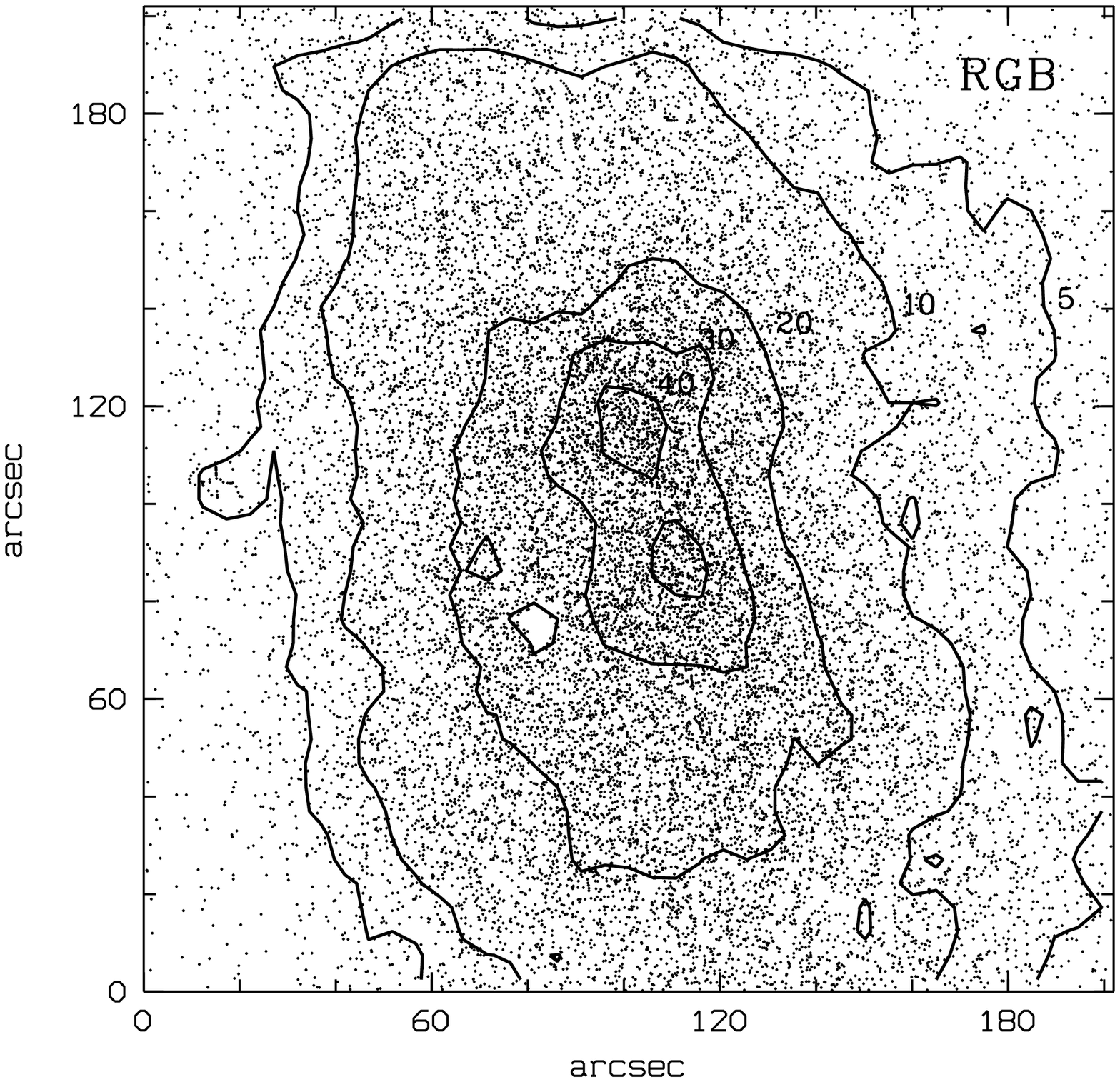}
\caption{Spatial distribution of the stellar populations in UGCA\,92.
The main sequence stars is bounded by $(V-I)_0\leq0$.
The boundaries of the shown blue loop are $0.0\leq(V-I)_0\leq0.5$.
RS stars correspond to the range $1.0\geq(V-I)_0\leq1.5$ and
$I_0\leq21.5$. The red giant population bounded by $(V-I)_0\geq0.5$ and
$I_0\geq23.0$. Stellar density contours of the populations are shown with the solid lines.}
\label{fig:space}
\end{figure*}

We have selected few stellar populations depending on their
position in the CM diagram. The result of this selection is presented in
Fig.~\ref{fig:space}. The upper main sequence stars (upper left panel) has
the unreddened colour $(V-I)_0\leq0$, their age range is about 10--100~Myr.
These youngest stars of the galaxy are mainly concentrated in few knots
which highlight regions of ongoing star formation and form actual irregular
structure of UGCA\,92.

In the upper right panel of the figure we show supposed blue-loop (BL) stars
selected by a colour $0.0\leq(V-I)_0\leq0.5$. Their age range is
$\sim$10--200~Myr. Spatial distribution of these stars
is considerably more smooth and regular in comparison with MS stars though they
are well-concentrated to the central parts of the galaxy.

A selection criterion for red supergiant stars
(RS), with ages in range $\sim$10~Myr -- 1~Gyr,
(lower left panel) was quite conservative ($1.0\leq(V-I)_0\leq1.5$ and
$I_0\leq21.5$) to avoid possible contamination of the spatial structure by young
and intermediate age AGB stars (which are redder) and RGB (which are fainter).
As a result, the selected population is not numerous and smoothly distributed
in the central part of the galaxy nearly along the major axis.

A most numerous population is the red giant branch ($>$1~Gyr) observed nearly
$(V-I)_0\geq0.5$ and $I_0\geq23.0$. The stars are widely distributed in
the image with apparent concentration to the centre of UGCA\,92 and nearly
the major axis of the object. The density of the stars is smoothly decayed to
the edge of the galaxy, which probably extends beyond the image boundary.

\section{Star formation and environment}

\begin{figure*}
\centerline{ \includegraphics[width=0.7\textwidth,clip]{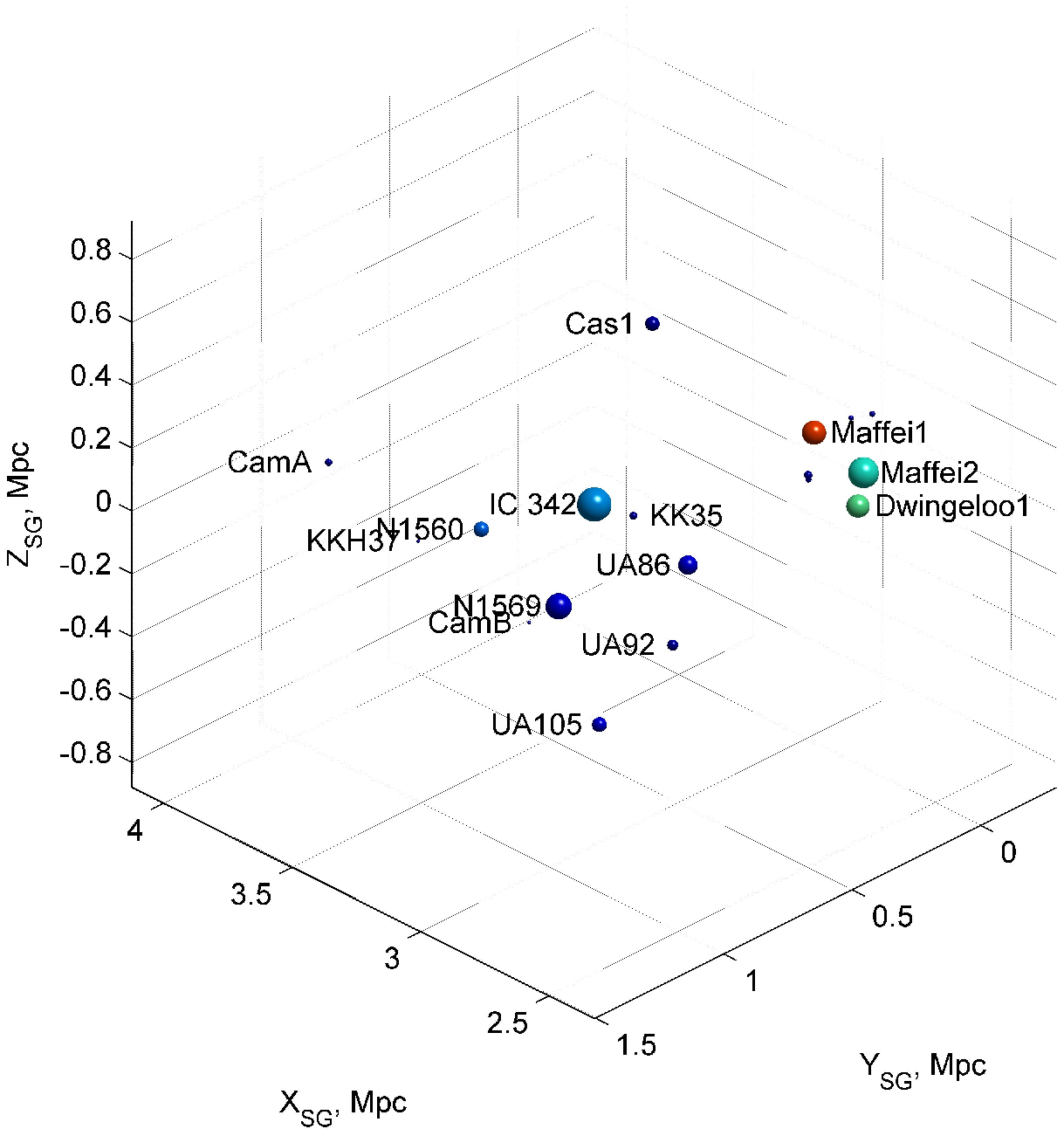} }
\caption{The 3D structure of the IC\,342 group. The size of the data cube is
$1.8\times1.8\times1.8$ Mpc. The giant spiral IC\,342 is placed in the centre
of the cube.}
\label{fig:struct}
\end{figure*}

Maffei -- IC\,342 is highly obscured nearby galaxy complex. 
These galaxies are gathered in two groups.
One group is around the giant
face-on galaxy IC\,342 and another one is around the pair of E+S galaxies Maffei\,1
and Maffei\,2. According to \citet{idk05}, the groups contain eight
members each. One should keep in mind, that the distances to the group members could
be highly uncertain due to the high uncertainty in the Galactic extinction 
in this direction. The IC\,342 group structure is shown in the
Fig.~\ref{fig:struct}. There are 19 galaxies in
the plot which situated at distances $\leq 1.8$ Mpc from IC\,342.
The distances were taken from the Catalog of Neighboring Galaxies \citep{CNG}.
The data on the distances were updated recently (Karachentsev, private
communications). The two galaxy groups, around IC\,342 and around Maffei\,1
and Maffei\,2 could be seen in the figure. Most of the complex members are
dwarf galaxies. The morphological types of the objects are coded by a colour
according to \citet{devau} and \citet{CNG} 
from the early types (red) to the late types (dark blue).
It is interesting to note that all the dwarf satellites of IC\,342, 
including the galaxy under study UGCA\,92 are irregulars. 
An absence of dwarf spheroidal
satellites subsystem in the group can imply that dSphs still were not discovered,
because their lower surface brightness and high Galactic extinction put serious
observational constrains.

The dwarf irregular galaxy under study UGCA\,92 is situated at a linear distance
$D=440$ kpc from the gravitational centre of the group IC\,342. The three
nearest to UGCA\,92 galaxies are UGCA\,86 (a linear distance $D=260$ kpc),
UGCA\,105 ($D=300$ kpc) and NGC\,1569 ($D=360$ kpc). UGCA\,92 is often
considered to be the companion to the starburst galaxy NGC\,1569. They
have been known to have close radial velocities, 
$V_{LG}=93$ \kms{} for UGCA\,92 \citep{figgs} and 
$V_{LG}=106$ \kms{} for NGC\,1569 \citep{things}. 
Indeed, except for the giant spiral IC\,342 ($D=320$ kpc from NGC\,1569) 
and a very small irregular galaxy Cam\,B ($D=190$ kpc from NGC\,1569), 
UGCA\,92 is the spatially closest companion to NGC\,1569.

A median radial velocity of galaxies within IC\,342 group is $V_{LG}=244$ \kms{} 
with the velocity dispersion of 79 \kms{}. 
The pair of galaxies NGC\,1569 -- UGCA\,92 has a maximal peculiar velocity within the group. 
The question arises whether these galaxies form a real subsystem within IC\,342 group.
Indeed, the uncertainty in UGCA\,92 distance is 350 kpc (see Tab.~\ref{t:param}),
which is pretty close to derived linear separation between NGC\,1569 and UGCA\,92.
The main source of this uncertainty is a huge Galactic absorption.
Taking into account the close radial velocities and close positions on the sky,
these galaxies could form a tight physical pair.
In this case we could expect to find a correlation in their star formation.
From the other hand, the estimation of the linear distance between the galaxies is reasonable.
Therefore, the similar velocities and positions on the sky are just coincidence in a virialized system.

NGC\,1569 is the well-known nearest strong starburst galaxy. \citet{vallenari} have
found evidence of a recent burst of star formation from about 15 to 4 Myr ago.
One more distinct burst was found not older than 150 Myr ago.
Later \citet{angeretti} have found three recent distinct starbursts using HST/NICMOS data: 
13 -- 37 Myr, 40 -- 150 Myr and $\sim$ 1 Gyr ago (for 2.2 Mpc distance). 
A presence of bright \hii{} regions also indicate substantial ongoing star formation.
The galaxy contains two extremely luminous super-star clusters \citep{hunter00}. 
In the last work 45 other young clusters also have been identified,
the most of them have an age $<30$~Myr.
The distance to this galaxy was uncertain for the long time. The accurate
TRGB distance ($3.36\pm0.20$ Mpc) was measured by \citet{grocholski} using
\textit{HST}/ACS images. Stellar content of the NGC\,1569 halo was studied in great detail
recently by \citet{rys11}. Judging from these two papers, the old
(about 10 Gyr) RGB stars, including the outer halo in this starburst galaxy
should be more metal reach ([Fe/H] $\simeq -1$) than we measured for the old
RGB stars of UGCA\,92. One of the interesting conclusions of \citet{rys11} is
that the outer stellar halo of the starburst galaxy NGC\,1569 is not tidally
truncated and it is not outward extension of the inner disk, but instead
it is the distinct stellar halo with no evident age/metallicity gradient, i.e.
the starburst phenomenon is highly centrally concentrated. Probably,
such a morphology could be the result of past interaction/merging.
The question arises whether we can find any signs of past interactions for
the dwarf galaxies using the information on star formation of these objects.

\begin{figure}
\centerline{\includegraphics[height=12cm]{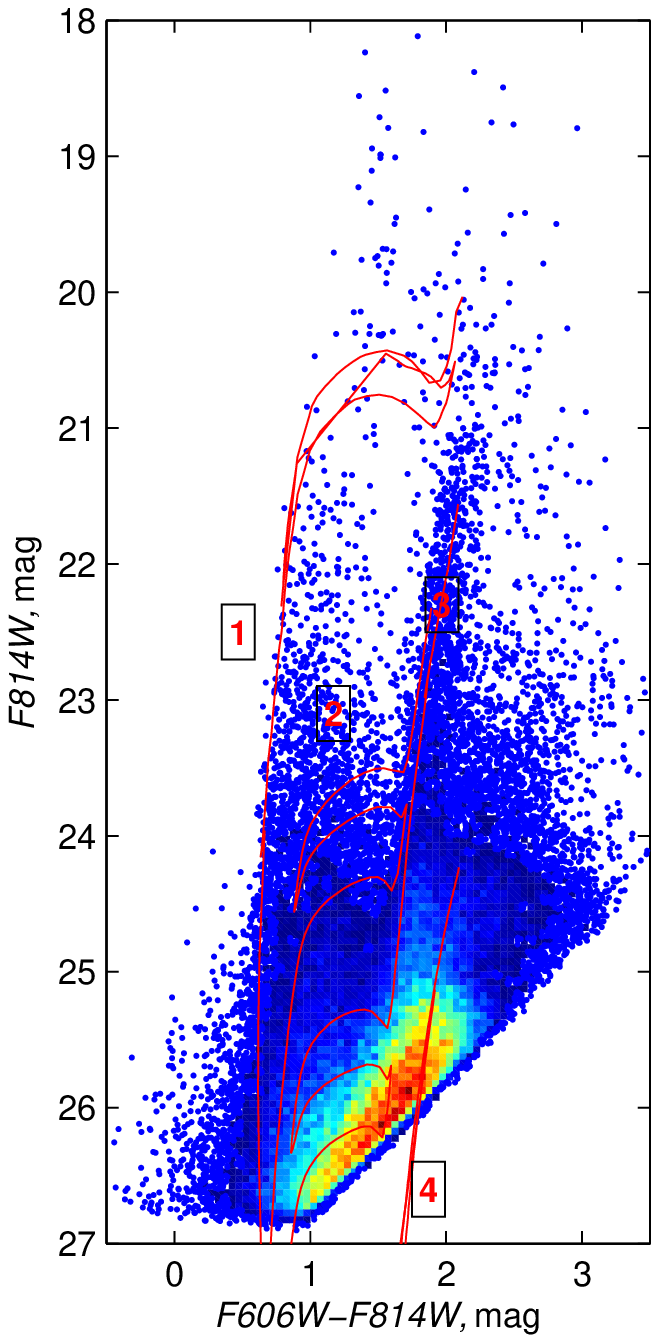}}
\caption{The $(\textit{F606W}-\textit{F814W})$, \textit{F814W} CMD of the dwarf galaxy UGCA\,86.
The magnitudes are not corrected for Galactic extinction.
The same Padova theoretical isochrones \citep{girardi00} as for UGCA\,92 CMD were shown.}
\label{fig:ua86cmd}
\end{figure}

Besides UGCA\,92, in the subgroup of the closest neighbours NGC\,1569--UGCA\,92--UGCA\,86--UGCA\,105 only UGCA\,86 has observations
deep enough (\textit{HST}/ACS images within the project number 9771, PI I. Karachentsev)
to judge about overall star formation from the photometry of resolved stars.
The colour-magnitude diagram of this galaxy is presented in
the Fig.~\ref{fig:ua86cmd}. The both UGCA\,92 (D = 3.03 Mpc) and UGCA\,86
(D = 2.96 Mpc, \citet{ketal06}) are at nearly the same distance from us. 
However, UGCA\,86 has the brighter
absolute stellar magnitude ($M_B = -17.95$) and larger angular size.
We have measured more resolved stars in the ACS field. 
An upper main sequence and helium-burning blue loop stars are found at $\textit{F606W}-\textit{F814W}<1.6$ mag. 
The well-populated RSG and the huge AGB are situated above the TRGB at $\textit{F606W}-\textit{F814W}>1.6$ mag and $\textit{F814W}<25.14$ mag. 
Galactic extinction for UGCA\,86 is even higher ($E(B-V)=0.94$ mag according to \citet{dustmap}). 
A photometric limit and the large colour excess seriously affect the RGB zone in the CMD (see Fig.~\ref{fig:ua86cmd}). 
This fact reduces seriously a reliability of a computational modelling of the star formation older than 1 Gyr.
Therefore, we only fitted a number of theoretical isochrones to the young stellar population of UGCA\,86.
The age and metallicity of these populations are similar in UGCA\,86 and UGCA\,92.

\hi{} and \ha{} observations of NGC\,1569, UGCA\,92 and UGCA\,86 were performed
by different authors. Numerous \ha{} knots were detected in all the three
galaxies, tracking the ongoing star formation events in these objects
\citep{hod95,kin98,kar2010}.
High-sensitivity \hi{} maps of NGC\,1569 show an evidence for a companion \hi{} cloud
connected with the galaxy by a low surface brightness \hi{} bridge. At the edge
of NGC\,1569 it coincides with \hi{} arcs \citep{stilisr98}.
The hydrogen cloud is apparently not correlated with any optical
satellites/counterparts.
\citet{stilisr02} argued, that about 10 per cent by mass of all \hi{} in NGC\,1569
have unusually high velocities. Some of this \hi{} may be associated with the mass
outflow evident from \ha{} measurements, but some may also be
associated with the NGC\,1569's \hi{} companion and \hi{} bridge, in which case, infall
rather than outflow might be the cause of the discrepant velocities.
\hi{} observations by \citet{stil05} show a complex structure of UGCA\,86, with two
separate components: a rotating disk and a highly elongated spur that is
kinematically disjunct from the disk.

Hereby, summarizing the cited results, we could conclude, that hydrogen
at the outskirts of the starburst galaxy NGC\,1569 is highly disturbed
with the signs of infall, but no similar features were detected in the
neighbouring dwarfs UGCA\,92 and UGCA\,86.

It should be note,
that all the data known to date could not give us the particular age, when
the series of the distinct intensive star bursts in NGC\,1569 was started.
\citet{angeretti} mention that the last starburst should occur
8 -- 27 Myr ago if the distance to the galaxy is 2.9 Mpc instead 13 -- 37 Myr
assuming the distance of 2.2 Mpc. The revised accurate distance to
NGC\,1569 is 3.36 Mpc according to \citet{grocholski}. It is possible that
other two starburst periods occurred 40 -- 150 Myr ago and about 1 Gyr ago will
be shifted to the younger ages assuming the new distance.
Our data on recent star formation in the companion UGCA\,92 galaxy
show substantial and continuous star formation in the last 500 Myr with the enhancement
at about 200 -- 300 Myr. Therefore, our results do not indicate a direct connection between 
the recent star formation in the two galaxies.

\section{Conclusions}
We have derived a quantitative star formation history of the dwarf irregular galaxy UGCA\,92
situated in the highly obscured nearby galaxy group IC\,342.
Due to a low Galactic latitude ($l = +10.5^{\circ}$) the extinction is very high ($E(B-V) = 0.79\pm0.13$ according to \citet{dustmap}) in this direction.
The star formation history was reconstructed using \textit{HST}/ACS images of
the galaxy and a resolved stellar population modelling. According to our measurements,
84 per\,cent of the total stellar mass were formed during the star formation
occurred about 12 -- 14 Gyr ago. A metallicity range of these stars is
${\rm [Fe/H]}=[-2.0:-1.5]$\,dex. There are also signs of marginal
star formation 4 -- 6 Gyr ago. A metallicity of these stars is similar
to the metal abundance of the oldest stellar population.

UGCA\,92 has a typical morphology of an irregular dwarf with apparent associations
of bright blue stars in its body. Numerous \hii{} knots were also detected in
the galaxy, tracking the ongoing star formation. 
According to our measurements, recent star formation was started about 1.5 -- 2 Gyr ago
and continuing to the present.
We modelled the star formation history with good time resolution for the recent
star formation event. The continuous star formation in this period shows
moderate enhancement from about 200 Myr to 300 Myr ago.
A mean star formation rate in the last 500 Myr is 
$4.3\pm0.6\times10^{-2}$ \msol{} yr$^{-1}$ and 
$3.9\pm0.6\times10^{-2}$ \msol{} yr$^{-1}$ in the last 50 Myr.
The mass portion of the stars formed in the last 500 Myr
is 7.6 per\,cent of the total mass of formed stars.
A metallicity of the recent star formation event is determining with
a large uncertainty due to relatively poor statistic in comparison with
sufficiently more numerous old stars. However, the measurements show that
a significant part of the young stars is evidently metal enriched.
It is very likely that the ongoing star formation has a metallicity of
$-0.6$ -- $-0.3$ dex.

UGCA\,92 is often considered to be the companion to the starburst
galaxy NGC\,1569. They have been known to have close radial velocities,
$V_{LG}=93$ \kms{} for UGCA\,92 \citep{figgs} and $V_{LG}=106$ \kms{} for NGC\,1569 \citep{things}.
The linear distance between the galaxies is $D=360$ kpc.
These two objects are evidently could be considered as a close pair of galaxies.

Another dwarf galaxy close to UGCA\,92 is UGCA\,86 with the linear distance $D=260$ kpc.
Our HST/ACS data allow to judge about overall star formation from the photometry of resolved stars. Theoretical isochrone fitting shows an apparent similarity of the resolved stellar
populations in the two galaxies. 

It is worth to note, that the mean metallicity of the old RGB stars measured by us in UGCA\,92 is lower, than the known metallicity of the halo RGB stars in NGC\,1569.  

Comparing our star formation history of UGCA\,92 with that of NGC\,1569
reveals no causal or temporal connection between recent star formation 
events in these two galaxies. We suggest,
that the starburst phenomenon in NGC\,1569 has not related to the closest
dwarf neighbours and does not affect their star formation history. 
 
Probably, detailed N-body modelling of the group within
300 kpc from IC\,342 is necessary to clarify a reason of recent star bursts
in NGC\,1569.

\section*{Acknowledgements}
The work was supported by the Russian Foundation for Basic Research (RFBR) grant 11--02--00639 and Russian-Ukrainian RFBR grant 11--02--90449. 
We acknowledge the support of the Ministry of Education and Science of the Russian Federation, 
the contract 14.740.11.0901.
This work was partially supported by Physical Sciences Department program
of Russian Academy of Sciences.
We are thankful to the anonymous referee for the very useful comments to the paper and to Scott Trager for his kind help with the text preparation. 
We acknowledge the usage of the HyperLEDA database (\url{http://leda.univ-lyon1.fr}).

\bibliographystyle{mn2e}
\bibliography{dwarfirr}   

\bsp
\label{lastpage}

\end{document}